\documentclass[10pt,conference]{IEEEtran}

\usepackage{cite}
\usepackage{amssymb}

\usepackage{amsmath,amsfonts}
\usepackage{algorithmic}
\usepackage{graphicx}
\usepackage{textcomp}
\usepackage{xcolor,colortbl}
\usepackage{outlines}
\usepackage{soul}
\usepackage{comment}
\usepackage[linesnumbered,ruled,vlined]{algorithm2e}
\usepackage{balance}
\usepackage{microtype}
\usepackage{multirow}
\usepackage[normalem]{ulem}
\useunder{\uline}{\ul}{}
\usepackage{booktabs}
\usepackage{hyperref}
\usepackage{todonotes}
\hypersetup{
    colorlinks=true,
    allcolors=blue,
}

\usepackage{tikz}

\def\BibTeX{{\rm B\kern-.05em{\sc i\kern-.025em b}\kern-.08em
    T\kern-.1667em\lower.7ex\hbox{E}\kern-.125emX}}
    
\usepackage{titlesec}
\titlespacing\section{0pt}{4pt plus 4pt minus 1pt}{0pt plus 2pt minus 1pt}
\titlespacing\subsection{0pt}{4pt plus 4pt minus 2pt}{0pt plus 2pt minus 1pt}
\titlespacing\subsubsection{0pt}{4pt plus 4pt minus 2pt}{0pt plus 2pt minus 1pt}

\begin{document}

\title{Architectural Archipelagos: Technical Debt\\in Long-Lived Software Research Platforms}

\author{
    \IEEEauthorblockN{Marcelo Schmitt Laser*, Duc Minh Le$^\dagger$, Joshua
    Garcia$^\ddagger$, and Nenad Medvidovi\'c*}
    \IEEEauthorblockA{University Of Southern California*, Bloomberg LP$^\dagger$, University of California, Irvine$^\ddagger$ \\
   \{schmittl@usc.edu, dle50@bloomberg.net, joshug4@uci.edu, neno@usc.edu\}\vspace{-4ex}}
}

\maketitle

\begin{abstract}
This paper identifies a model of software evolution that is prevalent in large, long-lived academic research tool suites (3L-ARTS). This model results in an ``archipelago'' of related but haphazardly organized architectural ``islands'', and inherently induces technical debt. We illustrate the archipelago model with examples from two 3L-ARTS archipelagos identified in literature.
\end{abstract}

\begin{IEEEkeywords}
Software research, proof-of-concept, archipelago
\vspace{-4ex}
\end{IEEEkeywords}

\section{Introduction}
\label{sec:intro}

In the academic world, small software proofs-of-concept (POC) are most frequently developed by researchers to run experiments that tend not to generalize. POC are usually discarded shortly after they fulfill their purpose (e.g., when a submitted paper is accepted, or M.S. thesis or Ph.D. dissertation completed). However, there are cases when research-based software grows into large, multi-purpose components, tools, frameworks, workbenches, and/or environments (e.g., when the developed capabilities are perceived by a new Ph.D. student as a good foundation for their own dissertation research). We refer to such systems as 3L-ARTS (\emph{L}arge, \emph{L}ong-\emph{L}ived \emph{A}cademic \emph{R}esearch \emph{T}ool \emph{S}uites). 3L-ARTS possess unique characteristics and suffer from unique complications: due to the haphazard processes by which such research systems emerge, they inherently accumulate technical debt. In turn, this directly hampers their transition to other research groups or to industrial usage, despite containing state-of-the-art technology~\cite{tilley2003challenges}. 

\looseness-1
Over the past three decades, we have experienced the challenges of developing, maintaining, and evolving 3L-ARTS in more than one instance. Collectively, these experiences have taken place at multiple universities and have involved dozens of graduate students guided by academic researchers, as well as collaborators from industry. We have repeatedly witnessed similar issues in trying to reuse third-party academic research tools. Through these experiences, we have identified a phenomenon, peculiar to academic research, in which independently developed extensions of a core pool of capabilities result in progressively increasing departures from the original architecture. 

In order to characterize and further study this phenomenon, identify its (dis)advantages, and design mitigation strategies for the rapid accumulation of technical debt resulting from it, we propose a conceptual model called \textit{software archipelago}. An archipelago consists of a set of ``islands'' (see Figure~\ref{fig:archipelago}). Each island is a software tool that has emerged as an outcrop of a previously existing island to address a set of closely related problems. Each individual island tends to exhibit different software architectural traits as it is developed by different people---most often, graduate students---for different purposes and used in different ways. Due to the inherent high coupling between the islands, a person attempting to use one of these islands typically needs to understand and comply with a large subset, if not all, of the archipelago's architecture. The archipelago is, in turn, likely to contain inconsistent or conflicting architectural decisions across its islands.

These characteristics result in 
complex software systems
: each added island 
risks creating direct or indirect couplings to existing islands. Eventually, archipelagos fall into disuse as students perceive that building new capabilities from scratch is easier than trying to refactor and reuse existing capabilities. 

This paper introduces the archipelago model, explains how and why it tends to emerge in academic-research settings, and illustrates it with examples from two real-world archipelagos. We discuss lessons-learned from both growing our own, and attempting to reuse other research groups' archipelagos. This discussion is intended to shed light on what is a relatively common, but mostly ignored phenomenon of software development in an academic setting, and to inspire future work on avoiding the introduction of and/or reducing the existing significant technical debt in research-off-the-shelf software.
\section{Motivation}
\label{sec:motiv}

The inherent difficulty in developing and managing 3L systems has held the attention of researchers for decades. Most such systems are in industry- or government-related domains, and have been the focus of a range of attempts at providing guidelines, useful patterns, and reference architectures (e.g.,~\cite{SHAW:1996, BASS:2003, TAYLOR:2009, durik2012, Deienbck2009ContinuousQC}). Open-source software systems have also recently received considerable attention, as they present examples with long, reliable, and detailed histories of development.

\looseness-1
In contrast, despite regularly facing significant degrees of architectural decay and technical debt, academia has yet to widely adopt practices to understand, control, and reduce these phenomena. This is unsurprising, given that academic research does not put primary emphasis on software development, instead encouraging the ``publish or perish'' model which tends to result in brittle proof-of-concept implementations and ``throwaway code''~\cite{FOOTE:1999}.
On the one hand, software engineering research is an applied discipline that frequently results in non-trivial tools. 
On the other hand, those tools often tend to be short-lived due to the nature of and forces shaping academic research. 

\looseness-1
Graduate students and post-doctoral researchers, who are the principal drivers of development in an academic setting, are \emph{temporary} project contributors whose primary goal is to advance a research idea and demonstrate its feasibility, rather than develop mature software tools. Similarly, funded projects are limited in duration and are most often evaluated on the novelty of the underlying idea as opposed to the sophistication of the resulting tool support. We note that the forces shaping the development and evolution of software tools may be similar in other areas of computer science, and likely beyond; however, this paper primarily draws on our experience with software engineering research and tools over the past three decades.

\looseness-1
The software engineering research community has recognized and tried to address this problem, resulting in several developments in recent years. One such development is the growing expectation that research claims be supported with extensive empirical evidence, which indirectly encourages more robust tools. This is reflected in 
recent initiatives toward \emph{open science} in software engineering~\cite{ROSE2019}\cite{ROSE2020}, and 
at recognizing research that is \emph{reusable} and \emph{replicable}~\cite{ACMBADGES}. 
It is not clear that these initiatives will directly impact researchers' long-standing motivations that inherently place their priorities elsewhere.\footnote{As a telling counter-example, at least one Distinguished Paper Award winner at ESEC/FSE 2020, at the time of this paper's submission the most recent major software engineering conference, released none of the underlying algorithms' implementations or evaluation datasets.}

Together, these forces have shaped the \emph{archipelago} model of software development prevalent especially in academic research. Section~\ref{sec:archi} describes the model and Section~\ref{sec:casestudy} illustrates it with publicly available data from two 3L-ARTS.
\section{Archipelago}
\label{sec:archi}

The archipelago model is conceptually depicted in Figure~\ref{fig:archipelago}. The original project, \emph{O}, represents the first major tool produced in a family (e.g., by one or more PhD students). 
This effort may result in 
a number of publications, while additional publications may require major or minor extensions to \emph{O}, designated with \emph{OM} and \emph{Om}, respectively. 
We provide a categorization of the extensions, driven by our experience and by reviewing several 3L-ARTS from literature. This is not intended as a complete characterization or as a rigorous mechanism for identifying the type and scope of a specific extension. Instead, it is a descriptive tool that helps 
understand the forces that introduce significant technical debt in a 3L-ARTS archipelago.

Major extensions to a project are typically introduced by new contributors (students or post-docs), likely after the developer(s) of \emph{O} have left the project.\footnote{In certain situations, the original developer may ``take \emph{O} with them'', resulting in a fork with two independent evolution paths. This is what occurred, e.g., with the c2.fw project discussed in Section~\ref{sec:casestudy}.} A major extension will result in the introduction of significant new functionality to solve a previously unsolved problem. It will often require non-trivial additions to \emph{O} and adaptations to its APIs. A 
heuristic that helps 
identify a major extension is that it tends to yield one or more major 
peer-reviewed publications, or an entire dissertation's worth of work. We identify two types of major extensions:

\begin{figure}[t]
    \centering
    \includegraphics[width=\columnwidth]{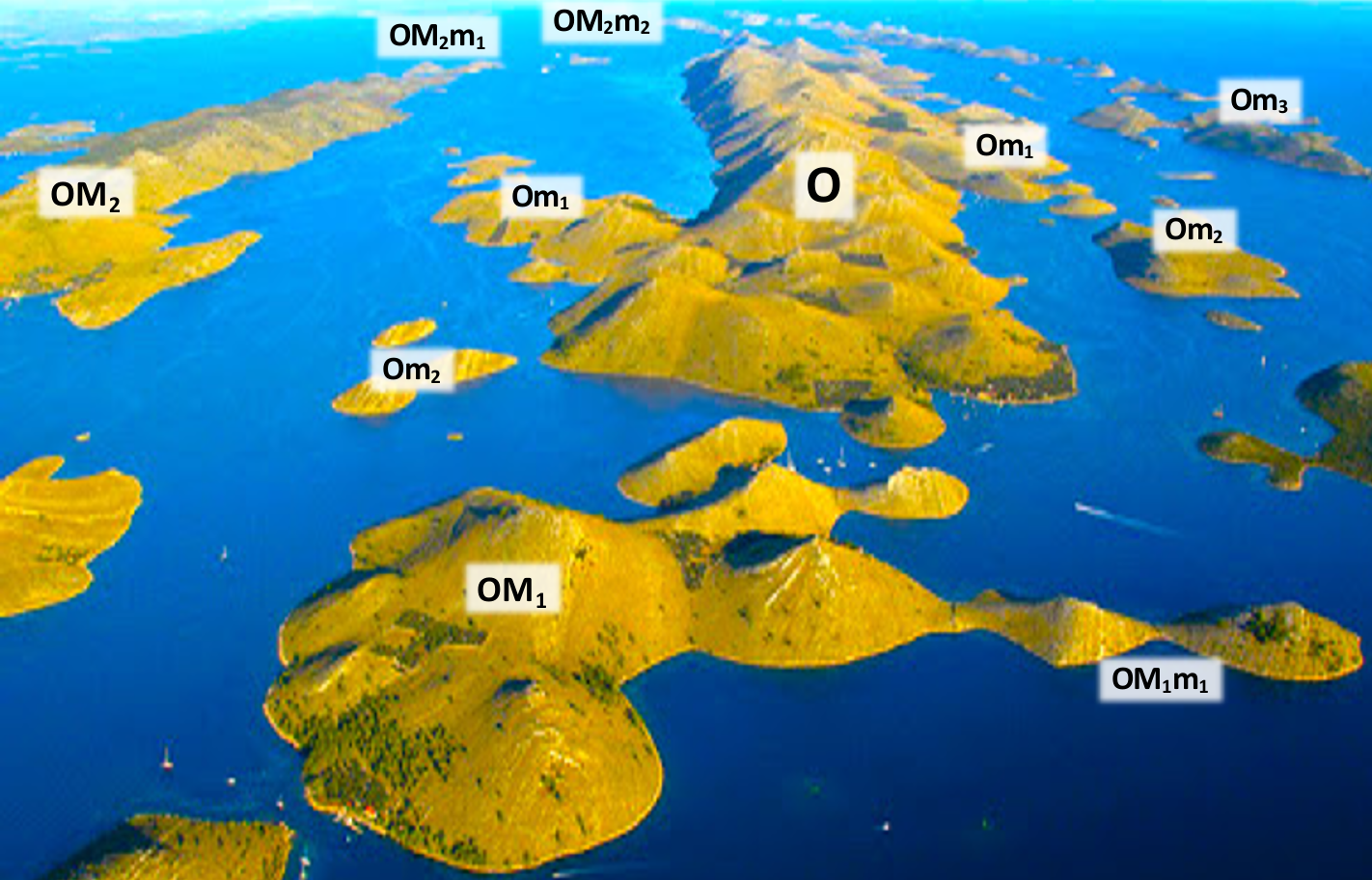}
    \caption{\looseness-1 The archipelago model of developing and evolving 3L-ARTS. \emph{O} designates the original project; \emph{M} designates a major extension to the original project; \emph{m} is a minor extension to the original project or to one of its major extensions.}
    \vspace{-2mm}
    \label{fig:archipelago}
\end{figure}

\begin{itemize}
\item An extension that results in significant new functionality and solves a previously unsolved problem, but is directly enabled by \emph{O} (e.g., by leveraging its APIs or built-in extension facilities), is of type \emph{OM$_{1}$}.
\item An extension that, in addition to the above characteristics of \emph{OM$_{1}$}, requires going beyond \emph{O}'s built-in extension mechanisms and adapting its APIs, is of type \emph{OM$_{2}$}.
\end{itemize}

\looseness-1
Minor extensions to a project may be introduced by the original developer(s) of \emph{O} or new contributors. A minor extension is characterized by slight-to-moderate departures from \emph{O}, such as by adapting existing or adding relatively simple new functionality, or modifying \emph{O} to support additional inputs/outputs or to combine it with third-party utilities. In externally-visible terms, a minor extension will typically result in no more than one publication: rather than aiming to answer a new research question, a minor extension attempts to generalize or improve the results of prior work. There are three types of minor extensions:

\begin{itemize}
\item The  extensions that are directly enabled by \emph{O} (e.g., by leveraging its APIs) are of type \emph{Om$_{1}$}.
\item The  extensions that require slight-to-moderate departures from \emph{O} (e.g., by slightly adapting or adding relatively simple functionality to \emph{O}) are of type \emph{Om$_{2}$}.
\item The extensions that are inspired by \emph{O}  but require more significant departures from it (e.g., modifying its functionality to support additional inputs/outputs and/or combining it with third-party utilities) are of type \emph{Om$_{3}$}.
\end{itemize}

Finally, it is possible to provide a minor extension to a major extension, much in the same manner in which the original project is extended. It may also be possible in principle to introduce a major extension to a major extension of \emph{O}. We identified two such candidate extensions in the c2.fw project family discussed in Section~\ref{sec:casestudy}. An interesting question that goes beyond this paper's scope is, at what point such extensions begin to depart from the point of origin \emph{O} so much that they should be treated as entirely new projects.

Systems that fit the archipelago model allow new contributors to quickly create extensions using their choice of tools and technologies: there are frequently no explicitly enforced architectural constraints aside from those encoded in the existing APIs. However, this also means that new contributors may neglect, or even violate, design decisions made by previous contributors. This leads to architectural dissimilarities between software islands that range from the code level, such as naming conventions and design patterns used, to the architectural level, such as architectural styles followed and connector types used. Extensions may even be created using different technology, such as a new programming language, with little consideration given to the complexity of managing such a project. All of this results in accumulation of technical debt as each island imposes its own set of issues.


\section{Archipelago case studies}
\label{sec:casestudy}

A number of 3L-ARTS archipelagos have been reported in literature. Just a few examples are Alloy~\cite{JACKSON:2002}, ArchStudio~\cite{DASHOFY:2007}, Titan~\cite{XIAO:2014}, and Archie~\cite{MIRAKHORLI:2014}. In this section, we highlight two archipelagos: the ARCADE family of projects developed between 2009 and present~\cite{GARCIA:2013,BEHNAMGHADER:2017,LASER:2020}, and the c2.fw family of projects developed between 1995 and 2015~\cite{TAYLOR:ICSE1995,taylor1996component,dashofy_dissertation,safi2015detecting}. Both families have publicly available code bases. More important to our analysis, both have had extensively documented architectures. We have composed the two systems' archipelagos based on their publication histories: we traced the topics covered in different papers, zeroed in on each paper's lead author, and subsequently tried to locate that author's Ph.D. dissertation or M.S. thesis, to confirm the postulated development timeline.


\subsection{Case Study 1: ARCADE}

ARCADE is an extensible workbench for supporting the recovery of software systems' architectures, and for evaluating architectural change and decay~\cite{LASER:2020}. It first arose out of the need to evaluate existing architecture-recovery techniques~\cite{GARCIA:2013} as part of a single student's PhD dissertation research~\cite{garcia2014unified}. A critical, unplanned aspect of constructing ARCADE's initial version was the need to re-implement several published architecture recovery techniques that did not have available working implementations. It is important to note that, at this early stage, ARCADE's authors reported no intent to turn it into a 3L-ARTS. This scenario illustrates a common software engineering research challenge: while the authors of each of the recovery techniques integrated into ARCADE fulfilled their individual objectives by having published their work, the slow decay or outright loss of those techniques' implementations became latent technical debt that is spread community-wide and that eventually had to be paid off by ARCADE's developers.

\looseness-1
The initial student-developer of ARCADE realized early on that he needed to provide several supporting utilities to achieve the project's evaluation goals. The set of components for software architecture recovery and the utilities allowing for their interplay together form the original island \emph{O} of ARCADE's archipelago. At this early point, ARCADE was a relatively compact workbench that resulted in at least one major publication~\cite{GARCIA:2013} and a Ph.D. dissertation~\cite{garcia2014unified}. 

As a system 
based on the dataflow architectural style~\cite{TAYLOR:2009}, ARCADE's analyses generated several artifacts that 
served as inputs for further extensions. Along with the 
small size of the original workbench, these characteristics led to a drive to extend ARCADE with new functionalities that would further its research value: detecting instances of architectural decay, visualizing different facets of software systems' architectures, predicting emergence of architectural issues, etc. 

Based on the information presented in ARCADE's initial publication~\cite{GARCIA:2013}, none of the above-mentioned extensions were originally considered. The first major ARCADE extension, of type \emph{OM$_{1}$}, emerged as a result of a second Ph.D. dissertation~\cite{ducsthesis}. It focused on detecting architectural decay caused by poorly managed dependencies among system components and poor separation of concerns. Several checks for the presence of such ``architectural smells''~\cite{GARCIA:2009A} had already been added to ARCADE's original implementation as a minor extension of type \emph{Om$_{1}$}. However, those checks were intended as a proof-of-concept and were neither comprehensive nor properly modularized, proving to be a direct source of unforeseen technical debt. The major extension provided a 
more comprehensive treatment of architectural decay, but in the process 
duplicated some 
existing functionality, as discussed in~\cite{LE:2018A}. Furthermore, our analysis of ARCADE's implementation repository~\cite{arcade_bitbucket} indicates that both the original, partial (\emph{Om$_{1}$}) and subsequent, more complete (\emph{OM$_{1}$}) decay-detection capabilities relied on the same third-party static analysis libraries, which were duplicated, introducing further technical debt into the system.

One particularly interesting utility was added to the ARCADE archipelago as a minor extension of type \emph{Om$_{2}$} after the addition of the original set of decay detectors (\emph{Om$_{1}$})
: ARCADE-Controller. This was a framework of adapters to allow different islands to connect and execute in tandem~\cite{BEHNAMGHADER:2017}. Here, the archipelago proved to be a hindrance: unlike 
typical commercial software development, the research tools comprising ARCADE are not mutually constrained in any way. While the islands themselves are independent from neighbors, any component attempting to connect them will necessarily become entangled with all of them, and therefore accumulate technical debt every time a connected island (initially, \emph{O} and \emph{Om$_{1}$}) is changed or a new island (e.g., \emph{OM$_{1}$}) is added. In addition, ARCADE-Controller was developed using shell script, which is suitable for executing remotely on multiple servers but hard to maintain and extend. Ultimately, ARCADE-Controller appears to have been removed from the more recent versions of ARCADE~\cite{LASER:2020}, likely because the cost of maintaining it outweighed its benefits.

Two further major extension islands (both of type \emph{OM$_{2}$}) were introduced in ARCADE to enable different architectural visualizations. Visualization tools are inherently driven by the particular needs of individual stakeholders, and ARCADE is no exception. The new islands were developed by two different students: one specifically to visualize the results of decay detection~\cite{ducsthesis} and the other to provide a novel visualization of architectural evolution~\cite{NAM:2018}. A third visualization was developed, apparently in parallel by ARCADE's original author, to serve as a simple utility for capturing the internal structure of the origin island \emph{O}. The scope of this visualization was much narrower, yielding a minor extension of type \emph{Om$_{3}$}.

\looseness-1
The archipelago model was instrumental in allowing the three contributors to work rapidly and independently, as each visualization tool is decoupled from its neighbors (although reliant on the rest of the archipelago), going so far as to be developed with different technologies. However, once again, this resulted in significant technical debt: these tools are constrained by varying architectural styles and use a variety of servers, runtime environments, and file formats, so that maintaining them is bound to involve significant duplication of effort.


\subsection{Case Study 2: c2.fw}

c2.fw is a family of research projects that comprise a 3L-ARTS developed over two decades. The project family originated with the description of C2, a new software architectural style for GUI-intensive systems, in 1995~\cite{TAYLOR:ICSE1995} and seems to have ended with the publication of DeVa, an analysis tool for event-based systems, in 2015~\cite{safi2015detecting}. These two bookends form two of several large islands that emerged from this work. Our review of the publications describing this project family identified at least the following major extensions of the originally reported C2 implementation~\cite{taylor1996component}. Given the length restrictions as well as the lack of certain details in publicly available sources, we will provide only their high-level overview:
\begin{itemize}
    \item The C2 architectural style~\cite{TAYLOR:ICSE1995} and initial implementation framework developed to demonstrate its features~\cite{taylor1996component} comprise the original island~\emph{O}.
    \item This early work was used as the foundation of two Ph.D. dissertations focusing, respectively, on architectural specification and analysis (\emph{OM$_{1}$})~\cite{neno_dissertation, neno:icse:1999} and runtime architectural adaptation (\emph{OM$_{2}$})~\cite{oreizy_dissertation, oreizy:icse:1998}).
    \item The lessons-learned in building the above two extensions subsequently resulted in two separate project threads, each of which yielded several Ph.D. dissertations in its own right: ArchStudio~\cite{DASHOFY:2007,dashofy_dissertation} and Prism~\cite{MikicRakic:Middlware2003,MALEK:2005}. 
\end{itemize}

\looseness-1
An argument can be made that both ArchStudio and Prism were such significant departures from C2 that they formed their own origin islands, rather than major extensions. First, their publications indicate that they were developed by two disjoint subsets of researchers that grew out of C2's original team. Second, both ArchStudio and Prism eschewed the original project's focus on the C2 style: ArchStudio introduced (1)~Myx~\cite{dashofy_dissertation}, a new architectural style, (2)~PACE~\cite{suryanarayana2004pace}, an architectural pattern targeting security, as well as (3)~xADL~\cite{khare2001xadl,dashofy2001highly}, a style-independent architecture description language; meanwhile, Prism was intended as a framework for supporting architectural (1)~implementation~\cite{midas, MEDVIDOVIC:Mobility2010}, (2)~deployment~\cite{MikicRakic:CD2002, MALEK:TSE2012}, and (3)~analysis~\cite{GARCIA:ESEC/FSE2013, safi2015detecting} across a range of styles and application domains. 

\looseness-1
Although the exact nomenclature---major extension vs. new origin island---is debatable, it is not critical to our discussion. To acknowledge this ambiguity, we will designate ArchStudio and Prism with type \emph{O(M)$_{2}$}. There is a clear indicator of  technical debt that accumulated, initially in C2 (i.e., \emph{O}) and subsequently in both \emph{O(M)$_{2}$} projects:  an undisputed system-family relationship can be established across these sub-projects, yet each of them, and more narrowly, their specific islands, were continually phased out and new, very similar functionality rebuilt. For example, the original C2 implementation framework~\cite{taylor1996component} was extended multiple times~\cite{medvidovic_wicsa_2002} before being discontinued and its underlying ideas independently re-implemented \emph{twice}, as part of the Myx~\cite{dashofy_dissertation} and Prism-MW~\cite{MALEK:2005} frameworks. As another example, C2's original architecture modeling language and support tools~\cite{Medvidovic:1996} were eventually scrapped and two separate re-implementations emerged independently: xADL~\cite{khare2001xadl,dashofy2001highly} and DRADLE~\cite{neno:icse:1999}. Because of how  3L-ARTS archipelagos emerge and evolve, c2.fw reflects instances of old capabilities not only being phased out, but being rebuilt multiple times without any larger plan, only for those new capabilities to be abandoned themselves.
\section{Conclusion and Lessons Learned}
\label{sec:lessons}

Our experience indicates 
research environments are uniquely prone to developing software archipelagos. Unlike commercial software, research software is developed by teams whose members often have different, potentially conflicting goals. Team members are also often trained in different technologies, and are expected to develop the entirety of their work on their own, being unable to rely on specialists for particular project aspects or phases of the development cycle. Components are almost always expected to be short-lived, and are developed with the intention of being used almost exclusively by their authors. Most importantly, the existing incentive structure of academic research does not place value on maintainability, and therefore encourages the accumulation of technical debt.

Archipelagos allow researchers to meet their immediate goals 
efficiently. Individual students and teams are able to work independently 
while extending each others' work effectively, up to a point. Tools 
are developed rapidly, and with little overhead imposed by needing to learn new technologies. 
Lastly, a problem area can be explored in multiple directions 
without the need to settle for a shared solution for all team members, thus allowing the exploration necessary in academic research.

Despite these advantages, the archipelago model is not well suited for long-lived software. Technical debt is accumulated rapidly, and the combined lack of (1) focus on development and (2) cohesion between team members  means it is extremely difficult to repay it. Eventually, the amount of technical debt exceeds the benefits of extending the archipelago, at which point the archipelago stagnates and is often discarded.

We posit that, while existing development processes and techniques are able to mitigate the technical debt caused by archipelagos to some extent, they are not suited as a complete solution. As the software engineering community has come to place greater value on research-based \emph{software}, it is important that this phenomenon be further studied and appropriate, targeted solutions be developed for it.

\section{Acknowledgements}

The authors wish to thank Sam Malek for several insightful discussions on this subject. This work was supported in part by awards CNS-1823262, CNS-1823246, CNS-1823074, CNS-1823354, CNS-1823214, CNS-1823177, CNS-1823074, CCF-1823177, OAC-1835292, CCF-1816594, and CCF-1717963 from the National Science Foundation and the U.S. Office of Naval Research under grant N00014-17-1-2896. We would also like to thank the anonymous reviewers for their valuable feedback, which helped us to improve this work. 

\clearpage
\balance
\bibliographystyle{IEEEtran}
\bibliography{arcade-practice}

\end{document}